\documentclass[twocolumn, graphics, floatfix, a4paper, aps, prl, superscriptaddress, longbibliography]{revtex4-1}
\usepackage{graphicx}
\usepackage{float}
\usepackage{bm}
\usepackage[usenames,dvipsnames]{xcolor}
\usepackage[colorlinks=true,urlcolor=blue,citecolor=blue,linkcolor=blue]{hyperref}
\usepackage{amssymb,ulem,amsmath}
\usepackage{longtable}

\newcommand{\qql}{\textquotedblleft}
\newcommand{\qqr}{\textquotedblright}
\newcommand{\vc}[1]{\bm{\mathrm{#1}}}

\begin{document}
\title{Flat band induced non-Fermi liquid behavior of multicomponent fermions}

\author{Pramod Kumar}
\affiliation{Department of Applied Physics, Aalto University, FI-00076 Aalto, Finland}

\author{Sebastiano Peotta}
\affiliation{Department of Applied Physics, Aalto University, FI-00076 Aalto, Finland}
\affiliation{Computational  Physics  Laboratory,  Physics  Unit, Faculty  of  Engineering  and  Natural Sciences, Tampere  University,  P.O.  Box  692,  FI-33014  Tampere,  Finland}
\affiliation{Helsinki  Institute  of  Physics  P.O.  Box  64,  FI-00014,  Finland}

\author{Yosuke Takasu}

\author{Yoshiro Takahashi}
\affiliation{Department of Physics, Graduate School of Science, Kyoto University, Kyoto 606-8502, Japan}

\author{P\"aivi T\"orm\"a}
\email{paivi.torma@aalto.fi}
\affiliation{Department of Applied Physics, Aalto University, FI-00076 Aalto, Finland}

%
\begin{abstract}

We investigate multicomponent fermions in a flat band and predict experimental signatures of non-Fermi liquid behavior. We use dynamical mean-field theory to obtain the density, double occupancy and entropy in a Lieb lattice for $\mathcal{N} = 2$ and $\mathcal{N} = 4$ components. We derive a mean-field scaling relation between the results for different values of  $\mathcal{N}$, and study its breakdown due to beyond-mean field effects. The predicted signatures occur at temperatures above the N\'eel temperature and persist in presence of a harmonic trapping potential, thus they are observable with current ultracold gas experiments.
\end{abstract}

\maketitle

As interaction effects are enhanced in a flat Bloch band, remarkable ordered phases such as flat band ferromagnetism~\cite{Mielke:1993} and superconductivity~\cite{Kopnin:2011,Heikkila:2011,Peotta2015,Julku:2016} have been predicted. Quasi-flat bands, whose bandwidth is comparable or smaller than the typical energy scale of interactions, seem to explain why the critical temperature of the superconducting state recently observed in magic-angle twisted bilayer graphene is large compared to the Fermi energy~\cite{Viewpoint2020,Hu:2019,Julku:2020,Xie:2019}. Also the normal states above the critical temperature of ordered phases are expected to be non-trivial: since a non-interacting flat band system does not have a Fermi surface and is an insulator at any filling, a Landau-Fermi liquid is generally not expected~\cite{Tovmasyan:2018,Hofmann:2019}.
	
The strange metal phase of copper-based superconductors (cuprates) is the most well known example of a non-Fermi liquid phase, and is still not fully understood~\cite{Varma:1989,Lee:2006,Hussey:2008}. The repulsive Fermi-Hubbard model on a square lattice is considered a minimal model for the cuprates and according to recent numerical studies the crossover from a metallic (Fermi liquid) state at weak coupling to an antiferromagnetic insulator at strong coupling occurs through an intermediate non-Fermi liquid~\cite{Siimkovic:2020}. Strange metal behaviour has been observed experimentally both on the square lattice Hubbard model realized with optical lattices~\cite{Brown:2019} and in twisted bilayer graphene~\cite{Cao:2020}. For the Hubbard model on lattice geometries other than the simple square lattice, the existence of a non-Fermi liquid normal state is currently much less investigated. In the case of the Lieb lattice -- a typical flat band model, dynamical mean-field theory studies have established that non-Fermi liquid behavior is present in an extended region of the phase diagram~\cite{Kumar2017,Kumar2019}.
	
Experimental evidence of non-Fermi liquid behavior on composite lattice geometries such as the Lieb lattice is at present lacking. Here, we identify and calculate those experimental signatures of a non-Fermi liquid that can be most directly probed in ultracold gas experiments on composite lattice geometries. Since spin-related order or correlations, for instance magnetic order and pairing, are typically induced by a flat band, it is of interest to ask what happens if one goes beyond the case of spin-$1/2$ fermions~\cite{Cazalilla_2014}.
Recently a degenerate gas of bosonic isotopes of Ytterbium has been loaded in a Lieb lattice~\cite{Taie2015}, and the same can be done with fermionic ones~\cite{Yb173}.
This would provide the quantum simulation of the Fermi-Hubbard model with $\mathcal{N} > 2$ spin components~\cite{Cazalilla_2014}, in a paradigm flat band system. We investigate the non-Fermi liquid normal state of the repulsive Hubbard model on a Lieb lattice for both $\mathcal{N} = 2$ and $\mathcal{N} = 4$ cases, and find a scaling relation between them. We predict that the non-Fermi liquid properties manifest in the sublattice-resolved double occupancy and entropy -- all quantities that can be observed in ultracold gas experiments~\cite{Fuchs2011,Torma2015,mazurenko2017cold,PhysRevX.7.031025}. In particular,  the sublattice-resolved double occupancy for Ytterbium atoms can be measured by combining the sublattice-mapping technique, already demonstrated for the Lieb lattice~\cite{Taie2015,Taie2020}, with photoassociation-induced atom loss~\cite{Sugawa2011}.
	
As the temperatures considered here are above the magnetically ordered phase, standard 
ultracold gas setups can be used to verify our predictions and to experimentally demonstrate the non-Fermi liquid nature of a flat band system for the first time. Flat band lattices have been demonstrated with ultracold gas setups~\cite{Taie2015,PhysRevLett.108.045305}, and the Lieb lattice has  been  recently realized  also in atomic scale artificial matter~\cite{Drost2017,Slot2017,Yan:2019} and photonic systems~\cite{PhysRevLett.114.245503,PhysRevLett.114.245504,Baboux2016,Whittaker2018,Goblot2019}.
Our calculations in the $\mathcal{N} = 2$ case are relevant for non-Fermi liquid physics in a variety of systems, while ultracold Ytterbium and Strontium gases in the electronic ground state $^1S_0$~\cite{Fukuhara2007,Taie2010,Pagano2014, DeSalvo2010,Stellmer2013} provide both the $\mathcal{N} = 2$ and $\mathcal{N} = 4$ cases and allow testing of the predicted scaling relation.  

\begin{figure}
\includegraphics[width=1.0\linewidth]{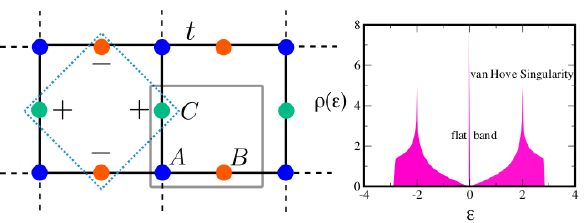}
\caption{\label{fig:lattice} 2D Lieb lattice: The convention for the unit cell (grey square box) and the labelling of the three sublattices ($\alpha = A,B,C$)  are shown.
The links represent hoppings with magnitude $t$. A localized state of the flat band is the linear superposition of the four sites of the $B$ and $C$ sublattices inside the dashed square box. The density of states $\rho(\varepsilon)$ of the noninteracting model is shown on the right.
}
\end{figure}

\begin{figure}[t]
\centering
\includegraphics[width=1.0\linewidth]{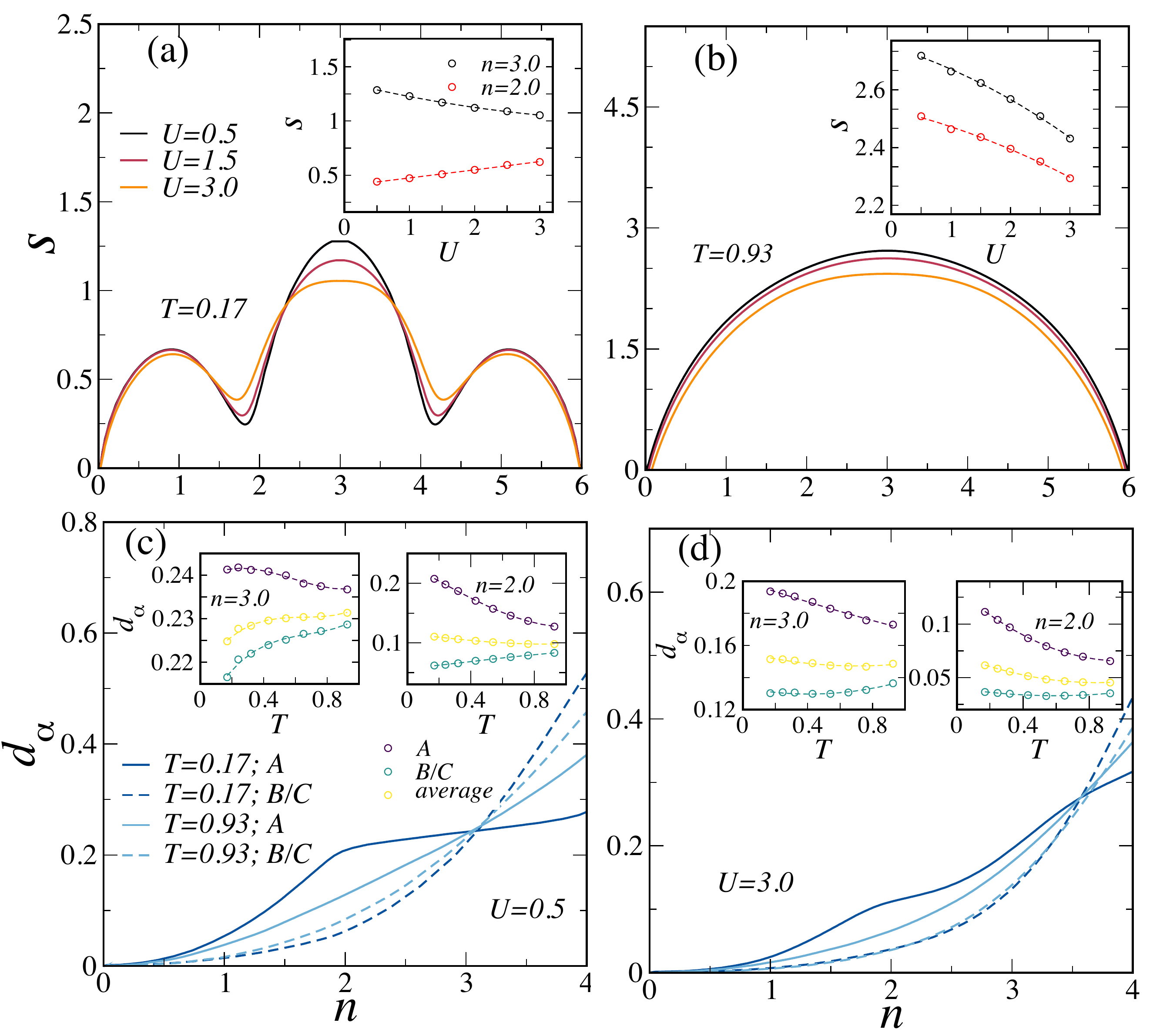}
\caption{\label{fig:one} \textbf{a)} Entropy per lattice site $s$ as a function of total filling $n = \sum_{\alpha} n_{\alpha}$ for $\mathcal{N} = 2$ component fermions at $T= 0.17$ and for three different values of the interaction strength. In the inset the entropy as a function of interaction strength is shown for filling $n = 3$ (half filling) and $n = 2$ (fully filled lowest band). \textbf{b)} Same as panel \textbf{a}, but at the different temperature $T= 0.92$. \textbf{c)} Sublattice resolved double occupancy $d_\alpha$ vs. filling $n$ at $U=0.5$ and for the same two values of temperature of panels \textbf{a} and \textbf{b}.
In the insets the double occupancy (both sublattice-resolved and averaged) vs. temperature at fixed fillings $n = 2,3$ is shown. \textbf{d}) Same as panel \textbf{c}, but for the different value of the interaction strength $U = 3$.}
\end{figure}

\noindent\textit{Model and methods} -- We consider in this work a Fermi-Hubbard model defined on the two-dimensional Lieb lattice model, which features a flat band at zero energy~\cite{Julku:2016}. The Lieb lattice model with nearest-neighbour hoppings and its density of states are shown in Fig.~\ref{fig:lattice}. The fermionic annihilation (creation) operator relative to sublattice $\alpha$ and unit cell $\vc{i} = (i_1,i_2)^T$ is $\hat{c}_{\vc{i}\alpha\sigma}$ ($\hat{c}^\dagger_{\vc{i}\alpha\sigma}$), while the occupation number operator is $\hat{n}_{\vc{i}\alpha\sigma} = \hat{c}_{\vc{i}\alpha\sigma}^\dagger\hat{c}_{\vc{i}\alpha\sigma}$.  The component (or \qql spin\qqr) index $\sigma$ labelling the fermionic operators takes values $\sigma = 1,\dots,\mathcal{N}$,  and we consider the case of four-component fermions (spin-$3/2$, $\mathcal{N}=4$) on top of the usual two (spin-$1/2$, $\mathcal{N}=2$). With this notation, the noninteracting Hamiltonian reads
\begin{equation}
\label{eq:Ham_nonint}
\mathcal{\hat H}_0 = \sum_{\vc{i},\vc{j},\alpha,\beta,\sigma} K_{\alpha,\beta}(\vc{i}-\vc{j}) \hat{c}^\dagger_{\vc{i}\alpha\sigma}\hat{c}_{\vc{j}\beta\sigma} -\mu \sum_{\vc{i},\alpha,\sigma}  \hat{n}_{\vc{i}\alpha\sigma}\,.
\end{equation}
Here $K_{\alpha,\beta}(\vc{i}-\vc{j})$ encodes the hopping matrix elements of magnitude $t$ between the nearest-neighbor sites in the Lieb lattice (see Fig.~\ref{fig:lattice} and Ref.~\cite{supplemental}). The hopping matrix is independent of the spin index $\sigma$, thus $\mathcal{\hat H}_0$ possesses an internal $SU(\mathcal{N})$ spin symmetry. In the following we use $t$ as the energy scale and set $t = \hbar = k_{\rm B} = 1$.
 
The full many-body Hamiltonian $\mathcal{\hat H} = \mathcal{\hat H}_0+\mathcal{\hat H}_{\rm int}$ is the sum of the noninteracting Hamiltonian and an interaction term $\mathcal{\hat H}_{\rm int}$, which is the generalization of the usual Hubbard interaction term for $\mathcal{N}\geq 2$ and preserves the $SU(\mathcal{N})$ symmetry as well
\begin{gather}
\hat{d}_{\vc{i}\alpha} = \sum_{{\sigma<\sigma'}}
\hat{n}_{\vc{i}\alpha\sigma}\hat{n}_{\vc{i}\alpha\sigma'}\,,\label{eq:d_def}\\
\label{eq:Hubbard_int}
\begin{split}
\mathcal{\hat{H}}_{\rm int} &= U\sum_{\vc{i},\alpha}\sum_{\sigma<\sigma'}\left(
\hat{n}_{\vc{i}\alpha\sigma}-\frac{1}{2}\right)\left(\hat{n}_{\vc{i}\alpha\sigma'}-\frac{1}{2}\right) \\
&=  U\sum_{\vc{i},\alpha}\hat{d}_{\vc{i}\alpha} -\frac{U(\mathcal{N}-1)}{2}\hat{N}+\text{const.}\
\end{split}
\end{gather}
The operator $\hat{d}_{\vc{i}\alpha}$ is the double occupancy operator for $\mathcal{N}$-component fermions and $\hat{N}=\sum_{\vc{i},\alpha,\sigma} \hat{n}_{\vc{i}\alpha\sigma}$ is the total particle number operator.  
We consider only the case of repulsive interactions ($U>0$).

In the following we focus on the expectation values of $d_{\vc{i}\alpha} = \langle \hat{d}_{\vc{i}\alpha}\rangle$ and $n_{\vc{i}\alpha} = \sum_{\sigma}\langle \hat{n}_{\vc{i}\alpha\sigma}\rangle$ as the main observables. We compute them using dynamical mean-field theory (DMFT)~\cite{RevModPhys.68.13} with the continuous time quantum Monte- carlo as impurity solver. The local density $n_{\alpha\sigma}$ is extracted from the local Green's function  ($n_{\alpha\sigma}=G_{\alpha\sigma}(\tau\rightarrow 0^-)$) computed with DMFT and the double occupancy at a given site  is evaluated from the Monte-Carlo perturbation order of the impurity solver~\cite{supplemental,PhysRevB.76.035116}. Since discrete translational invariance is assumed to be unbroken, the expectation values $n_{\vc{i}\alpha}$ and $d_{\vc{i}\alpha}$ do not depend on the unit cell index $\vc{i}$, which is dropped for simplicity. The filling of the lattice is defined as $n = \sum_{\alpha} n_\alpha$, thus $n = 3\mathcal{N}/2$ corresponds to half-filling for $\mathcal{N}$-component fermions. If the $SU(\mathcal{N})$ spin symmetry is also unbroken, that is no magnetic ordering occurs, the expectation values are also independent of the spin index, that is
$\langle\hat{n}_{\vc{i}\alpha\sigma}\rangle = \langle\hat{n}_{\vc{i}\alpha\sigma'}\rangle =  n_{\vc{i}\alpha}/\mathcal{N}$  for all $\sigma,\,\sigma'$ and
$\langle\hat{n}_{\vc{i}\alpha\sigma_1}\hat{n}_{\vc{i}\alpha\sigma_2}\rangle = \langle\hat{n}_{\vc{i}\alpha\sigma_3}\hat{n}_{\vc{i}\alpha\sigma_4}\rangle = d_{\vc{i}\alpha}/\binom{\mathcal{N}}{2}$ for all $\sigma_1\neq \sigma_2,\,\sigma_3\neq \sigma_4$. 
At  the temperatures considered here  magnetic ordering does not occur according to our results and the above conditions are always satisfied. Another observable of interest is the entropy per lattice site, $s$, which has been measured in many optical lattice experiments~\cite{Schneider1520,PhysRevX.7.031025}. The entropy is obtained from the occupation number using the relation $s(\mu,U,T)=\frac{1}{3}\int^{\mu}_{-\infty}\partial_{T} n(\mu,U,T)d\mu$~\cite{supplemental}.

\noindent\textit{Entropy and double occupancy} -- If $f(T,n,U)$ denotes the free energy per lattice site, one has $s = -\partial_Tf$ while the derivative with respect to the coupling constant $U$ gives the double occupancy averaged over the unit cell $d = \frac{1}{3}\sum_{\alpha}d_\alpha = \partial_U f$. Thus one has the Maxwell's relation
\begin{equation}
\label{eq:Maxwell}
\frac{\partial s}{\partial U} = -\frac{\partial d}{\partial T}\,.  
\end{equation}
Note that there is no sensible way to separate the entropy into contributions associated to single sublattices, while this makes sense for the double occupancy.

It is known in the case of the square and cubic lattices that, for $U$ sufficiently small, the double occupancy is a decreasing function of $T$ when temperature is increased from $T = 0$~\cite{Werner:2005,Fuchs2011}. This somewhat counterintuitive behavior is due to the fact that in a localized state the spin degrees of freedom store more entropy compared to a Landau-Fermi liquid, thus the system reduces the double occupancy and becomes more localized  upon heating. 

The Maxwell's relation~(\ref{eq:Maxwell}) provides an alternative explanation. The entropy in a Landau-Fermi liquid is linearly proportional both to the temperature and to the quasiparticle effective mass, $s\propto m_{\rm eff}T$, and since the effective mass generally increases with the coupling constant $\partial_U m_{\rm eff} >0$ for repulsive interactions~\cite{Yip:2014}, one has from Eq.~(\ref{eq:Maxwell}) that $\partial_T d < 0$. This argument holds only up to the temperature/energy scale at which the quasiparticles in the Landau-Fermi liquid are well defined, called the quasiparticle coherence scale $T_{\rm F}^*$~\cite{Werner:2005}. Above this scale the double occupancy increases with temperature due to  thermally-generated doubly occupied sites.
The initial decrease of double occupancy at low temperature occurs rather generally in Landau-Fermi liquids. One important example is liquid helium-3 where this effect is the root of Pomeranchuk cooling~\cite{Werner:2005,Richardson:1997}.
Pomeranchuk cooling has been demonstrated for $\mathcal{N}=6$ component fermions loaded in a cubic lattice~\cite{Taie:2012}.

\noindent\textit{${\mathcal{N} = 2}$ components} -- As shown in Fig.~\ref{fig:one}\textbf{a} the behavior of the entropy as a function of the coupling constant $U$ changes qualitatively depending on the filling. The triple peak structure of the entropy in Fig.~\ref{fig:one}\textbf{a} is a consequence of the density of states of the Lieb lattice, Fig.~\ref{fig:lattice}. The interesting observation is that, for fillings close to $n = 2$ one has $\partial_U s >0$, on the other hand for $n = 3$ the entropy decreases with $U$. The opposite behavior of the entropy at the two fillings $n = 2$ and $n = 3$ is emphasised in the inset Fig.~\ref{fig:one}\textbf{a}. For higher temperatures  (Fig.~\ref{fig:one}\textbf{b}) the entropy is always a decreasing function of $U$ at any filling.

The behavior of the double occupancy is consistent with that of the entropy as dictated by the Maxwell's relation~(\ref{eq:Maxwell}). As shown in the insets of Fig.~\ref{fig:one}\textbf{c}-\textbf{d} the average double occupancy is a monotonically increasing function of temperature for a half filled flat band ($n=3$), while it is decreasing at filling $n=2$. Moreover, the behavior of the sublattice-resolved double occupancy $d_\alpha$ depends qualitatively on the sublattice. We see from Figs.~\ref{fig:one}\textbf{c}-\textbf{d} that on sublattice $A$ the double occupancy decreases with temperature ($\partial_Td_A <0$), while on sublattices $B/C$ the behavior is opposite ($\partial_T d_{B/C} >0$). This striking difference is observed in the whole temperature range $0.15 < T < 1$ considered in Fig.~\ref{fig:one} and is particularly evident at half filling $n = 3$. This temperature range is above the magnetic phase, which occurs at around $T = 0.05$ for $U = 2$ according to our DMFT simulations, and just below the quasiparticle coherence scale $T_{\rm F}^*$ at which the double occupancy on the $A$ sublattice takes its minimum value.

\begin{figure}[t]
\centering
	\includegraphics[width=1.0\linewidth]{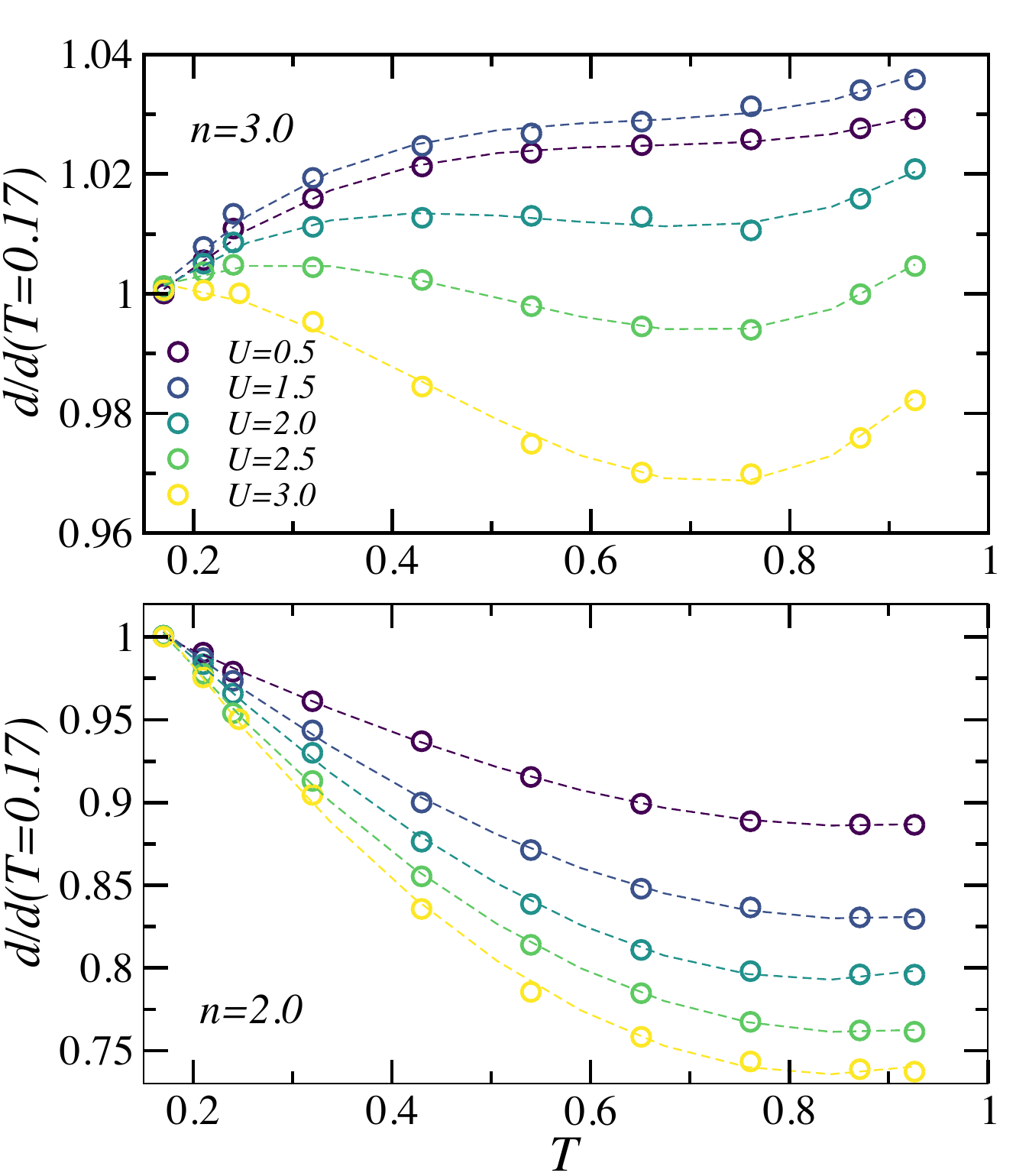}
	\caption{\label{fig:two}  Average double occupancy $d(T,n,U)$ rescaled by its value at the lowest temperature considered in this work $d(T = 0.17, n, U)$ as a function of temperature for different values of $U$  and for filling $n=3$ (top panel) and $n=2$ (bottom panel). The data for $U=0.5$ and $3$ are the same as the ones shown in the insets in Fig.~\ref{fig:one}\textbf{c}-\textbf{d}.}
\end{figure}

The interpretation of the results shown in Fig.~\ref{fig:one} is that the flat band is responsible for the non-Fermi liquid behavior ($\partial_U s = -\partial_T d < 0$). Indeed, the sublattice-resolved double occupancy provides the most compelling argument in this sense. The non-Fermi liquid behavior manifests only  in the double occupancy of sublattices $B/C$, where the flat band states have their support (see Fig.~\ref{fig:lattice}), while the double occupancy in the $A$ sublattice has the same behavior as,  for instance,  in a cubic lattice in the same temperature range~\cite{Werner:2005}. 
The flat band-induced non-Fermi liquid behavior can be observed for not too large interaction strength and modestly low temperatures.
As shown in Fig.~\ref{fig:one}\textbf{b} the entropy decreases with $U$ for all fillings at high temperatures, but this is a different effect, incoherent in nature, in which the flat band plays no role, and is observed also in simple square and cubic lattices.

Fig.~\ref{fig:two} illustrates how the flat band-induced non-Fermi liquid behavior is eventually  destroyed for large interaction strength. In Fig.~\ref{fig:two} the average double  occupancy, which does not resolve the different sublattices, is shown rescaled by its value at the lowest temperature considered here, that is $d(T,n,U)/d(T=0.17,n,U)$, to ease the visual comparison. From the top panel of Fig.~\ref{fig:two}, one can see that the average double occupancy at $n = 3$ is a monotonically increasing function of temperature for $U = 0.5$ and $1.5$, while a local minimum starts to develop at $U = 2$, and becomes well visible for $U = 2.5$ and $3$. This crossover from non-Fermi liquid to Fermi liquid behavior occurs because interactions tend to average out the contributions from all of the bands.
On the other hand, the standard Landau-Fermi liquid behavior always dominates in the averaged double occupancy away from half filling, as shown on the bottom panel of Fig.~\ref{fig:two}. Indeed,  at filling $n = 2$ the averaged double occupancy is always a decreasing function of temperature up to $T^*_{\rm F} \approx 0.9$, at which it attains its minimum as in the case of the cubic lattice~\cite{Fuchs2011}. 
  
\begin{figure}
  \includegraphics[width=1.0\linewidth]{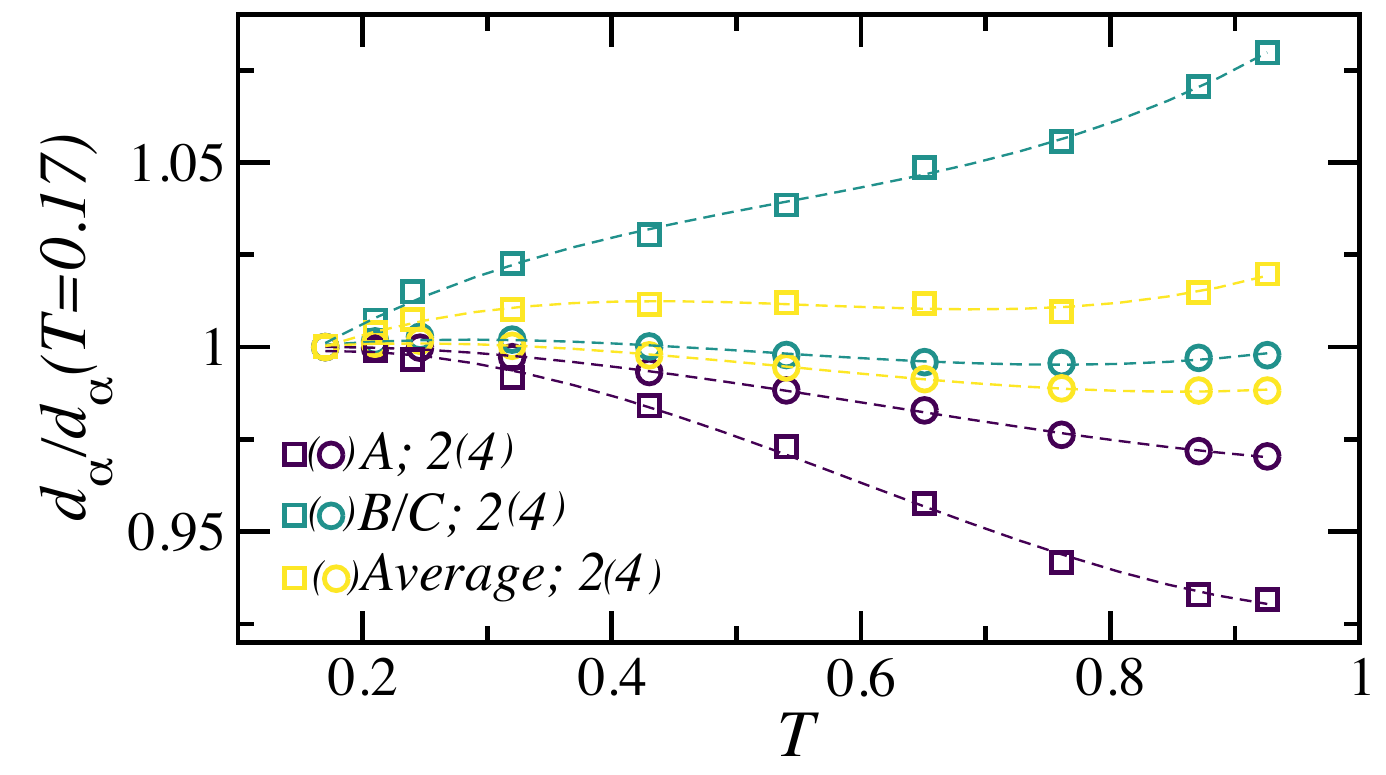}
  \caption{\label{fig:three} Sublattice-resolved and average double occupancy at half filling and $U=2$ rescaled by their value at the lowest temperature ($T_{\rm min} = 0.17$), namely $d_\alpha(T,n,U)/d_\alpha(T_{\rm min},n,U)$ and  $d(T,n,U)/d(T_{\rm min},n,U)$, respectively.} 
\end{figure}

\begin{figure}
\includegraphics[width=1.0\linewidth]{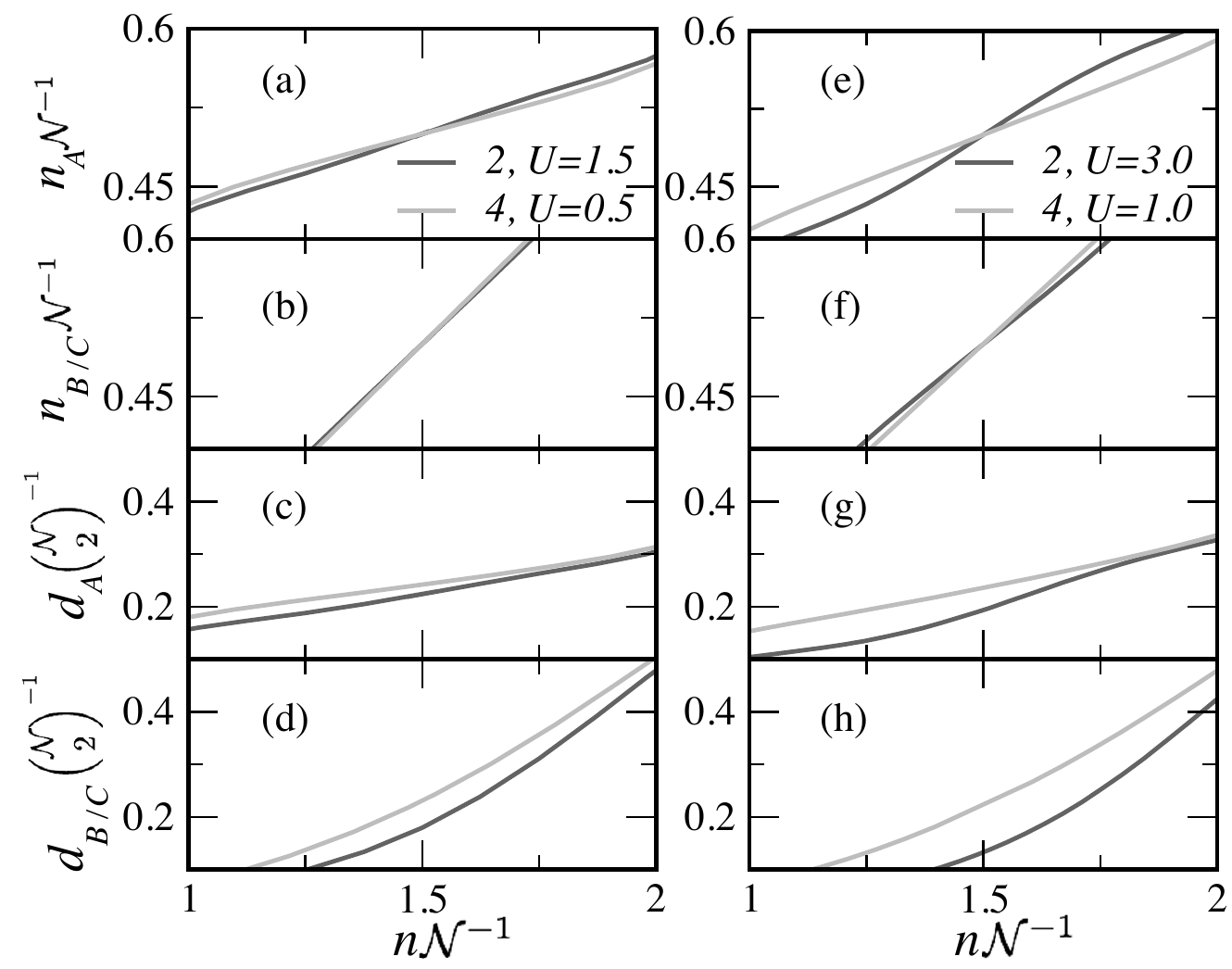}
\caption{\label{fig:four}  Comparison of DMFT results between pairs of values $(U,\mathcal{N})$ satisfying the scaling relation~(\ref{eq:scaling_rel}). On the left (right) column the black lines are results for the sublattice-resolved occupation number $n_{\alpha}$ and double occupancy $d_\alpha$ for $U = 1.5$ ($U = 3$) and $\mathcal{N} = 2$, while the grey lines are results for $U' = 0.5$ ($U' = 1$) and $\mathcal{N}' = 4$. The occupation number and the double occupancy are normalized in such a way that the curves for different number of components would coincide if the mean-field approximation was exact.}
\end{figure}

\noindent\textit{${\mathcal{N} = 4}$ components} -- In Fig.~\ref{fig:three} we compare the cases of $\mathcal{N}=2$ and $\mathcal{N}=4$ component fermions. 
For $\mathcal{N}=2$ components and $U = 2$  the behaviour of the averaged double occupancy is in the crossover region in between a Landau-Fermi liquid and a non-Fermi liquid, as discussed above in relation to Fig.~\ref{fig:two}. On the contrary, as shown in Fig.~\ref{fig:three}, for $\mathcal{N} = 4$ components the averaged double occupancy as a function of temperature at the same value of $U$ looks more like that of a Landau-Fermi liquid with the characteristic minimum at $T \approx 0.9$. We note also that the variation of the double occupancy with temperature is reduced for $\mathcal{N} = 4$ compared to $\mathcal{N}=2$, and that the double occupancy in the $B/C$ sublattices does not increase monotonically as it does for $\mathcal{N} = 2$. Apparently, the flat band-induced non-Fermi liquid behavior  disappears as the number of components is increased.

Here we propose a scaling argument to understand the results of Fig.~\ref{fig:three}. At the mean field level it is possible to show~\cite{supplemental} that the solution of the problem for a given pair of parameters $(U,\mathcal{N})$ provides also the solution for all pair of values $(U',\mathcal{N}')$ which satisfy the scaling relation 
\begin{equation}
\label{eq:scaling_rel}
U(\mathcal{N}-1) = U'(\mathcal{N}'-1)\,.
\end{equation} 
Inserting $U' = 2$, $\mathcal{N}' = 4$ and $\mathcal{N} = 2$ in Eq.~(\ref{eq:scaling_rel}), gives $U = 6$. In other words the result for $\mathcal{N}' = 4$ components in Fig.~\ref{fig:three} can be equivalently understood as the double occupancy of a model with $\mathcal{N} = 2$ and $U = 6$. At this large value of the coupling strength, one expects the non-Fermi liquid behavior induced by the flat band to be suppressed, and this is indeed the case as one can see from the data for $\mathcal{N}=4$  in Fig.~\ref{fig:three}.

In order to check the validity of the mean field approximation underlying the scaling relation~(\ref{eq:scaling_rel}), we compare DMFT results for pairs of parameters $(U,\mathcal{N})$ which satisfy the scaling relation, see Fig.~\ref{fig:four}.
One can observe that there is better agreement between the results for $\mathcal{N} = 2$ and $\mathcal{N}'=4$ for the lower values of the coupling constants (left column) with respect to the higher ones (right column). Indeed, one expects the mean field approximation to be accurate in the weakly interacting regime. Moreover a larger deviation is seen in the case of the double occupancy compared to the occupation number. This is also expected since the double occupancy is a quantity which is more sensitive to beyond mean-field correlations, indeed at the mean-field level it is simply the product of the occupation numbers ($\langle\hat{n}_{\vc{i}\alpha\sigma}\hat{n}_{\vc{i}\alpha\sigma'}\rangle = \langle\hat{n}_{\vc{i}\alpha\sigma}\rangle  \langle\hat{n}_{\vc{i}\alpha\sigma'}\rangle$ for $\sigma\neq \sigma'$). From Fig.~\ref{fig:four} one can conclude that the scaling relation holds qualitatively in the range of couplings of interest here ($0 \leq U \lesssim 3$ for $\mathcal{N}=2$).

\noindent\textit{Trap effects} -- Combined with the lattice potential, a harmonic trap is often used in ultracold gas experiments to confine the atoms. In Ref.~\cite{supplemental} we perform a DMFT study that takes into account such trapping effects via the local density approximation. Importantly, our prediction that the double occupancy in the Lieb lattice behaves in a qualitative different way depending on the sublattice (Fig.~\ref{fig:one}\textbf{c}-\textbf{d}) remains valid in presence of a harmonic trap. 

\noindent\textit{Conclusions} -- 
We identified signatures of non-Fermi liquid behavior in the entropy and double occupancy for a Lieb lattice system of $\mathcal{N} = 2$ and $\mathcal{N} = 4$ component fermions. We showed that the nonmonotonic behavior of the double occupancy, the fingerprint of a Landau-Fermi liquid, is not present at all for sufficiently small interactions. This is a consequence of the presence of a flat band in the band structure of the Lieb lattice. Indeed the non-Fermi liquid behavior in the double occupancy can be observed only in the sublattices on which the flat band states have their support, the $B/C$ sublattices, while on the $A$ sublattice the conventional behavior is observed.
Using mean field arguments, we derived a scaling relation~(\ref{eq:scaling_rel}) to describe the results for different numbers of components $\mathcal{N}$. 
The adequacy of the mean field approximation was investigated by means of DMFT and the scaling relation was found to be qualitatively correct in the range of couplings used in the present work. It will be interesting to probe the validity of this scaling relation in experiments as a direct indicator of beyond-mean field effects. Our predictions can be tested in currently available experimental ultracold gas setups, even in the presence of a harmonic trap: temperatures above the critical one are sufficient and only sublattice-resolved, not spatially resolved, imaging of the double occupancy is required. Our results open the route for experimental investigations and understanding of non-Fermi liquid normal state properties caused by flat band singularities.

\acknowledgements
This work was supported by the Academy of Finland under project
numbers 303351, 307419, 327293, 318987 (QuantERA project RouTe), by the Grant-in-Aid for Scientific Research of the Ministry of Education, Culture Sports, Science, and Technology / Japan Society for the Promotion of Science (MEXT/JSPS KAKENHI, Nos. JP17H06138, JP18H05405, and JP18H05228), Japan Science and Technology Agency CREST (No. JPMJCR1673), and by MEXT Quantum Leap Flagship Program (MEXT Q-LEAP, No. JPMXS0118069021).

\bibliography{biblio_capital}

\end{document}


\title{Supplementary material for \qql Flat band induced non-Fermi liquid behaviour of multicomponent fermions\qqr}

\author{Pramod Kumar}
\affiliation{Department of Applied Physics, Aalto University, FI-00076 Aalto, Finland}

\author{Sebastiano Peotta}
\affiliation{Department of Applied Physics, Aalto University, FI-00076 Aalto, Finland}
\affiliation{Computational  Physics  Laboratory,  Physics  Unit, Faculty  of  Engineering  and  Natural Sciences, Tampere  University,  P.O.  Box  692,  FI-33014  Tampere,  Finland}
\affiliation{Helsinki  Institute  of  Physics  P.O.  Box  64,  FI-00014,  Finland}

\author{Yosuke Takasu}

\author{Yoshiro Takahashi}
\affiliation{Department of Physics, Graduate School of Science, Kyoto University, Kyoto 606-8502, Japan}

\author{P\"aivi T\"orm\"a}
\email{paivi.torma@aalto.fi}
\affiliation{Department of Applied Physics, Aalto University, FI-00076 Aalto, Finland}

\maketitle

\begin{figure}
\includegraphics[scale=0.9]{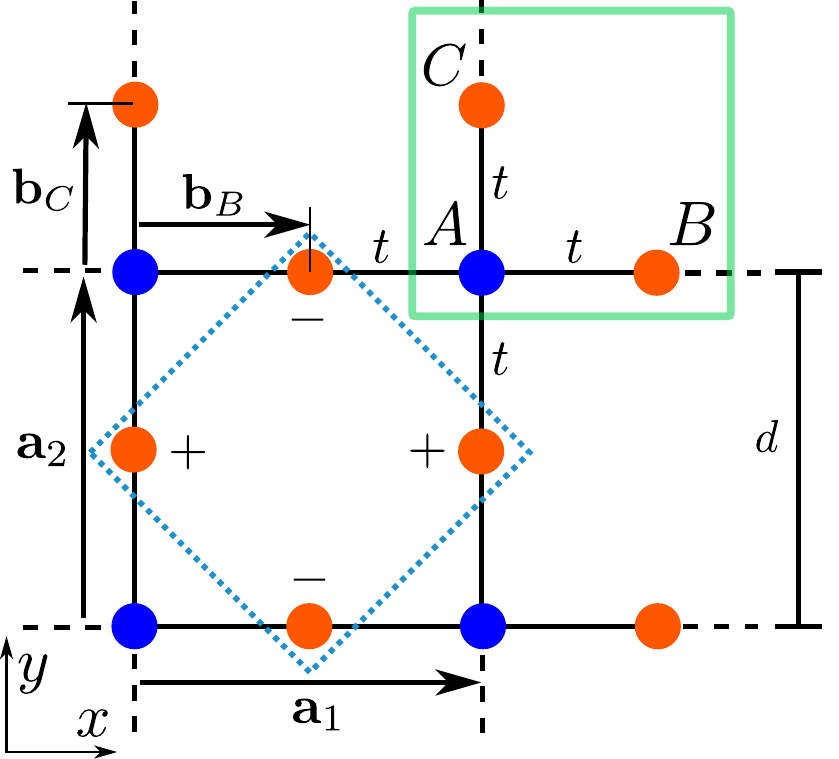}
\caption{\label{fig:lattice} Tight-binding model of the two dimensional Lieb lattice. The lattice spacing of the two-dimensional square Bravais lattice is denoted by $d$. Our convention for the unit cell (solid square box), the labelling of the three sublattices ($\alpha = A,B,C$) and the fundamental vectors of the Bravais lattice ($\vc{a}_1 = (d, 0)^T$, $\vc{a}_2 = (0, d)^T$) are shown. The vectors $\vc{b}_B = (d/2,0)^T$ and $\vc{b}_C = (0,d/2)^T$ defining the geometry of the sublattice structure  are also shown ($\vc{b}_A = \vc{0}$ for the $A$ sublattice).
The links represent nearest-neighbor hopping matrix elements with magnitude $t$. A characteristic localized state of the Lieb lattice, belonging to the flat band subspace, is also shown (dashed square box).}
\end{figure}

\section{Hopping matrix of the Lieb lattice}

For completeness we specify here the form of the hopping matrix $K(\vc{i}-\vc{j})$ that enters the definition of the noninteracting Hamiltonian $\mathcal{\hat H}_0$, Eq.~(1) in the main text. 
We denote by $\vc{r}_{\vc{j}} = j_1\vc{a}_1+j_2\vc{a}_2$ the lattice vector of the Bravais lattice associated to the unit cell labelled by $\vc{j}=(j_1,j_2)^T$.
With this notation, the Fourier transform $\widetilde{K}(\vc{k}) = \sum_{\vc{j}}e^{-i\vc{k}\cdot\vc{r}_{\vc{j}}}K(\vc{j})$ of the hopping matrix of the two-dimensional Lieb lattice with nearest-neighbour hoppings all equal to $t$ (Fig.~\ref{fig:lattice}), is given by 
\begin{equation}
\begin{split}
\widetilde{K}(\vc{k}) = t\begin{pmatrix}
0 & 1 + e^{ik_1}& 1 + e^{ik_2} \\ 
1 + e^{-ik_1} & 0 & 0 \\
1 + e^{-ik_2} & 0 & 0
\end{pmatrix}\,.
\end{split}
\end{equation} 
Here we have defined $k_i = \vc{k}\cdot \vc{a}_i$, with $\vc{a}_i$ the fundamental vectors of the Bravais lattice (Fig.~\ref{fig:lattice}), and the $3\times 3$ matrix $K(\vc{i})$ is composed of the matrix elements $K_{\alpha,\beta}(\vc{i})$ which encode the hoppings from sublattice $\beta$ to $\alpha$. As in the main text, we use the hopping amplitude $t$ as the energy scale and set $t = \hbar = k_{\rm B} = 1$.

\section{Dynamical mean-field theory for Lieb lattice}

To obtain all the results shown in the main text, we use dynamical mean-field theory (DMFT)~\cite{RevModPhys.68.13}  for the Hubbard model in the thermodynamic limit with a continuous time quantum Monte-Carlo impurity solver~\cite{PhysRevB.76.035116}. 
The essential approximation underlying DMFT is that of a local self-energy. This means in our case that the self-energy is independent of quasimomentum $\vc{k}$ and diagonal with respect to the sublattice index $\alpha$~\cite{RevModPhys.68.13}. Indeed, the local Green's function matrix for component $\sigma$   is given, within the DMFT approximation, by
\begin{equation}
  \mathbf{G}^\sigma(i\omega_n)= \frac{1}{N_{k}}\sum_{\vc{k}} \left( \mathbf{G}_{\mathbf{k} \sigma}^0(i\omega_n)^{-1}-\mathbf{\Sigma}^\sigma(i\omega_n) \right)^{-1}, \label{eq:dmft}
  \end{equation}
where $\mathbf{G}^0_{\vc{k}}(i\omega_n)^{-1}= (i\omega_n+\mu)\mathbf{1}-\widetilde{K}(\mathbf{k})$ is the noninteracting Green's function and  $\mu$ the chemical potential. The self-energy is assumed to be diagonal in the sublattice index, even though it can be different for different sublattices, i.e. $[\mathbf{\Sigma}^\sigma(i\omega_n)]_{\alpha,\beta} = \Sigma^{\sigma}_{\alpha}(i\omega_n)\delta_{\alpha,\beta}$~\cite{Kumar2019}.

For each sublattice $\alpha$ there is an effective single impurity Anderson model, which is defined by the dynamical Weiss mean-field
\begin{equation}
\mathcal{G}_{\alpha}^{\sigma}(i\omega_n)^{-1}=[\mathbf{G}^{\sigma}(i\omega_n)^{-1}]_{\alpha\alpha}+\Sigma^{\sigma}_{\alpha}(i\omega_n).\label{eq:weiss}
\end{equation}
Given the Weiss function $\mathcal{G}_{\alpha}^{\sigma}(i\omega_n)$ for all $\alpha$, we calculate the self-energy of each of the impurity problems using the continuous time quantum Monte-Carlo impurity solver. We use interaction expansion algorithm (CT-INT) for the impurity solver~\cite{PhysRevB.76.035116}. These new self-energies are then used again in Eq.~(\ref{eq:dmft}) and the process is iterated until a converged solution is found.

The density is computed from the one particle Green's function for different values of the chemical potential, Hubbard interaction and temperature. To calculate the entropy we use the following result~\cite{Schneider1520,PhysRevLett.105.065301} 
\begin{equation}
 s(\mu,U,T)=\frac{1}{3}\int^{\mu}_{-\infty}\frac{\partial n(\mu,U,T)}{\partial T}d\mu\,,\label{eq:entropy}
\end{equation}
which is derived from the Maxwell's relation $\frac{1}{3}\partial_T n(\mu,U,T) = \partial_\mu s(\mu,U,T)$ between the density (also called filling, see main text) $n(\mu,T,U)$ and the entropy  $s(\mu,T,U)$ per lattice site. We need to compute the entropy only for negative values of the chemical potential, $\mu < 0$, since  the particle-hole symmetry of the Lieb lattice implies $s(-\mu)=s(\mu)$. 
To calculate numerically the partial derivatives, we interpolate the results of a linear grid of chemical potential and temperature values for a given Hubbard interaction strength $U$.

The double occupancy at a given sublattice of the unit cell for a system with $\mathcal{N}$ components is obtained from the Monte-Carlo perturbation order
\begin{align}
\langle k \rangle_{\text{MC}}^\alpha&=-\frac{ U}{T}\sum_{\sigma<\sigma^\prime}\bigg\langle\bigg(\hat{n}_{\vc{i}\alpha\sigma}-\frac{1}{2}\bigg)\bigg( \hat{n}_{\vc{i}\alpha\sigma^\prime}-\frac{1}{2}\bigg)-\delta^2\bigg\rangle\,,
\end{align}
where $\delta$  is the impurity solver parameter chosen to be small for the half-filled case~\cite{PhysRevB.76.035116}. In the above equation, $U$ and $T$ are the Hubbard interaction and the temperature, respectively. From the perturbation order one can compute the double occupancy as
\begin{equation}
d_{\alpha}= (\mathcal{N}-1)\sum_{\sigma}\frac{ n_{\alpha \sigma}}{2} +\binom{\mathcal{N}}{2}\bigg(\frac{1}{4}-\delta^2\bigg)-\frac{T \langle k \rangle_{\text{MC}}^\alpha}{ U}\,, \label{eq12}
\end{equation}
where $d_\alpha=\sum_{\sigma <\sigma'}\langle \hat{n}_{\vc{i}\alpha\sigma} \hat{n}_{\vc{i}\alpha\sigma^{\prime}} \rangle$. If the $SU(\mathcal{N})$ symmetry remains unbroken, $n_{\alpha\sigma}=n_{\alpha\sigma^{\prime}}=n_{\alpha}/\mathcal{N}$, the above expression can be further simplified
\begin{equation}
d_{\alpha}= (\mathcal{N}-1)\frac{n_{\alpha}}{2} +\binom{\mathcal{N}}{2}(\frac{1}{4}-\delta^2)-\frac{T \langle k \rangle_{\text{MC}}^\alpha}{ U}\,. \label{eq12}
\end{equation}

\begin{figure}
\includegraphics[width=1.00\linewidth]{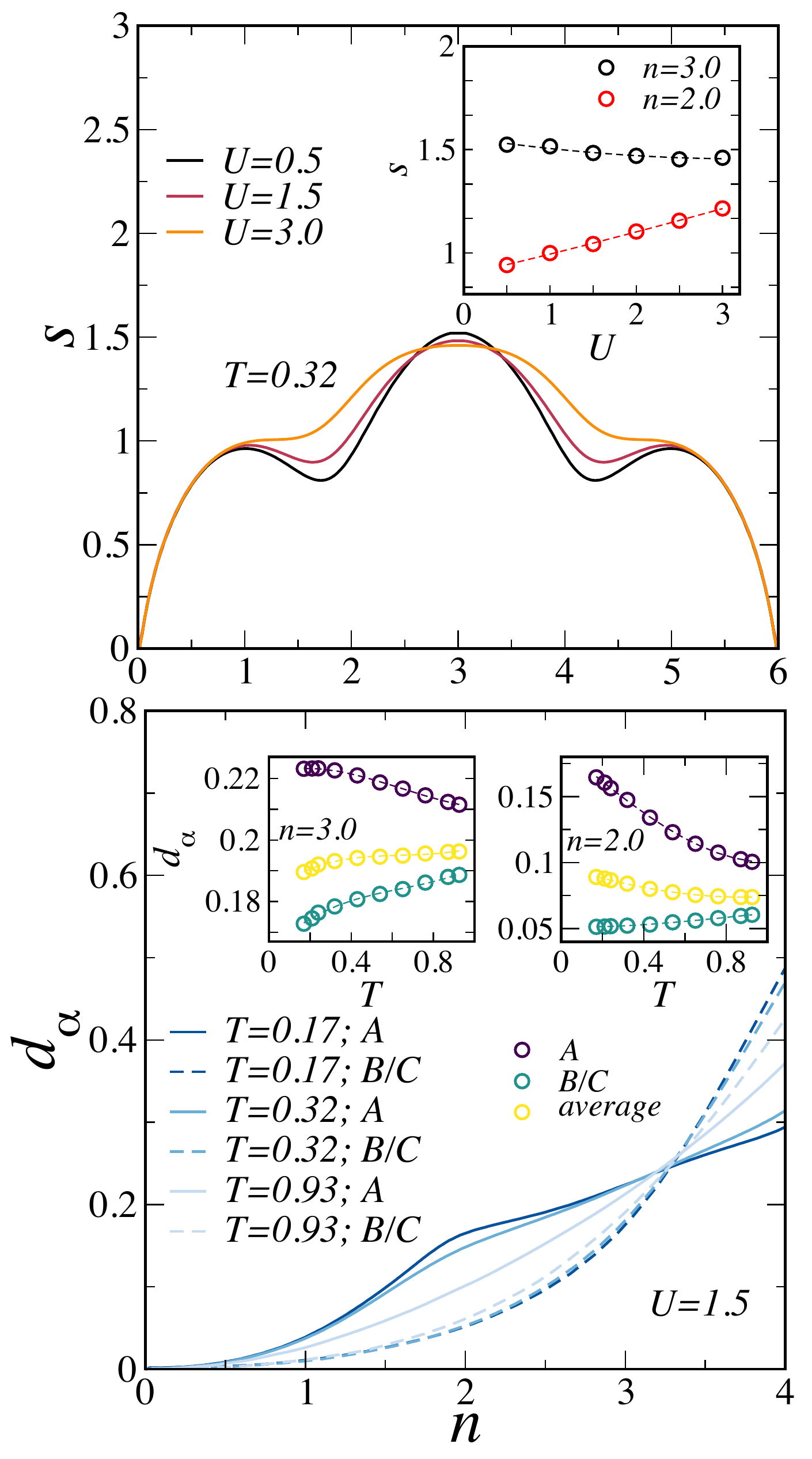}
\caption{\label{fig:two} Upper panel: entropy per lattice site $s$ as a function of total filling $n = \sum_{\alpha} n_{\alpha}$ for $\mathcal{N} = 2$ component fermions at $T= 0.32$ and for three different values of the interaction strength. In the inset the entropy as a function of interaction strength is shown for filling $n = 3$ (half filling) and $n = 2$ (fully filled lowest band). Lower panel: sublattice resolved double occupancy $d_\alpha$ vs. filling $n$ at $U=1.5$. In the insets the double occupancy (both sublattice-resolved and averaged) vs. temperature at fixed fillings $n = 2,3$ is shown.}
\end{figure}

\section{Entropy and double occupancy for intermediate temperature and interaction}

In Fig.~\ref{fig:two} we provide additional data which complement the one in Fig.~2 in the main text. In the upper panel of Fig.~\ref{fig:two}, the entropy per lattice site as a function of filling $n$ is shown at the temperature $T = 0.32$. By comparison with Fig.~2\textbf{a} in the main text, one can see that the window of fillings, for which the flat band-induced non-Fermi liquid behavior can be observed, is reduced with respect to the lower temperature $T = 0.17$, in favor of the standard Landau-Fermi liquid behavior (entropy increasing with increasing $U$). Moreover, the variation of the entropy at half-filling is also smaller at the higher temperature shown in the upper panel of Fig.~\ref{fig:two}, compared to Fig.~2\textbf{a} in the main text.

In the lower panel of Fig.~\ref{fig:two}, the double occupancy vs. filling is shown for $U = 1.5$, a value of the interaction strength which is in between the ones of panels \textbf{c} and \textbf{d} of Fig.~2 in the main text. Also for this value of $U$, the opposite behavior in inequivalent sublattices of the sublattice-resolved double occupancy as a function of temperature is clearly visible. Note also how the averaged double occupancy increases with temperature at filling $n = 3$, while it decreases at filling $n=2$ (see insets). Therefore the same behavior of the averaged double occupancy as shown in the insets of Fig.~2\textbf{c} in the main text, which depends in a qualitative way on the filling level, can be observed up to $U = 1.5$.

\section{Scaling with the number of components}

We provide here the derivation of the scaling relation of Eq.~(5) in the main text, which is based on the mean-field approximation. The mean-field approximation amounts to the replacement of the quartic interaction term of Eq.~(3) in the main text by a quadratic term which involves averages calculated over the mean-field ground state itself
\begin{equation}
\label{eq:mf_Hubbard_Ham}
\mathcal{\hat{H}}_{\rm int} \approx  U\sum_{\vc{i},\alpha}\sum_{\sigma<\sigma'}\left[\left(
\langle\hat{n}_{\vc{i}\alpha\sigma}\rangle-\frac{1}{2}\right)\hat{n}_{\vc{i}\alpha\sigma'} + (\sigma \leftrightarrow \sigma') \right]\,.
\end{equation}
Above the critical temperature for magnetic ordering, so that the $SU(\mathcal{N})$ symmetry is unbroken, the occupation number at a given site is the same for all components, that is $n_{\alpha\sigma}=n_{\alpha\sigma^{\prime}}=n_{\alpha}/\mathcal{N}$ for all $\sigma,\,\sigma'$. As a consequence one is reduced to diagonalize self-consistently the mean-field Hamiltonian given by
\begin{equation}
\label{eq:eff_mf_Ham}
\mathcal{\hat{H}}_{\rm mf,\sigma} = \mathcal{\hat{H}}_{\rm kin,\sigma} +U(\mathcal{N}-1)\sum_{\vc{i},\alpha}\left(
\langle\hat{n}_{\vc{i}\alpha\sigma}\rangle-\frac{1}{2}\right)\hat{n}_{\vc{i}\alpha\sigma}\,.
\end{equation}
Here $\mathcal{\hat{H}}_{\rm kin,\sigma}$ is the kinetic term for the $\sigma$ component, therefore the mean-field Hamiltonian is block diagonal in component space. Moreover, one needs to  diagonalize self-consistently the mean-field Hamiltonian $\mathcal{\hat{H}}_{\rm mf,\sigma}$ only for a single component since this automatically provides the solution for all other components. Note that the coupling constant $U$ and the number of components appear only as the combination $U(\mathcal{N}-1)$ in Eq.~(\ref{eq:eff_mf_Ham}). This means that the self-consistent solution of Eq.~(\ref{eq:eff_mf_Ham}) provides also the solution for all other pairs of values $(U',\mathcal{N}')$ which satisfy the scaling relation
$U(\mathcal{N}-1) = U'(\mathcal{N}'-1)$, as given in Eq.~(5) of the main text.

\section{Effect of the harmonic trapping}

The analysis presented here and in the main text has been restricted to the case of a translationally invariant lattice model in the thermodynamic limit. On  the other hand, it is important to take into account the effect of the harmonic trap, which is used to confine the atoms in experiments. Whereas it is nowadays possible to resolve, even in situ, the particle and spin density profiles, we want to show that it is not necessary to do so in order to test experimentally our prediction that the double occupancy in the Lieb lattice behaves in a qualitatively different way depending on the sublattice, as shown in Fig.~2 in the main text.

To this end, we focus on the sublattice-resolved and spatially-averaged double occupancy, that is the observable defined by
\begin{equation}
D_\alpha = \sum_{\vc{i}} \langle \hat{d}_{\vc{i}\alpha}\rangle\,.
\label{eq:D_a}
\end{equation} 
We first compute with DMFT the sublattice-resolved double occupancy in the thermodynamic limit for a dense grid of values of the chemical potential. Then $D_\alpha$ is obtained within the local density approximation, by assuming that the double occupancy at a given lattice site is equal to the one in the thermodynamic limit at the local chemical potential  $\mu-V(\vc{r}_{\vc{i}\alpha})$. The external potential  $V(\vc{r}_{\vc{i}\alpha})$ appears as an additional term in the noninteracting Hamiltonian
\begin{equation}
\label{eq:Ham_nonint}
\begin{split}
\mathcal{\hat H}_0 = \sum_{\vc{i},\vc{j},\alpha,\beta,\sigma} K_{\alpha,\beta}(\vc{i}-\vc{j}) \hat{c}^\dagger_{\vc{i}\alpha\sigma}\hat{c}_{\vc{j}\beta\sigma} 
\\ \quad+ \sum_{\vc{i},\alpha,\sigma}(V(\vc{r}_{\vc{i}\alpha})-\mu)  \hat{n}_{\vc{i}\alpha\sigma}\,.
\end{split}
\end{equation}
The position vector of the site in unit cell $\vc{i} = (i_1,i_2,i_3)^T$ and sublattice $\alpha$ is denoted by $\vc{r}_{\vc{i}\alpha} = i_1\vc{a}_1+ i_2\vc{a}_2 + i_3\vc{a}_3 + \vc{b}_\alpha$, where $\vc{b}_\alpha$ are the vectors that define the sublattice structure and are given in Fig.~\ref{fig:lattice}. 
Since we are considering the case of a stack of uncoupled Lieb lattices on parallel planes, we have introduced a third fundamental vector $\vc{a}_3 = (0,0,c)^T$ of the now three-dimensional Bravais lattice. The third fundamental vector $\vc{a}_3$ is perpendicular to $\vc{a}_1$ and $\vc{a}_2$. 
We provide results only for a single experimentally realistic instance of the external confining potential $V(\vc{r})$ given by
\begin{equation}
V(x,y,z) = \Omega_{11}\left(\frac{x}{d}\right)^2 +2\Omega_{12}\frac{x}{d}\frac{z}{c} + 
\Omega_{22}\left(\frac{z}{c}\right)^2 + \Omega_3 \left(\frac{y}{d}\right)^2\,.
\end{equation}
Here $d$ is the in-plane lattice constant (see Fig.~\ref{fig:lattice}), while $c$ is the distance between parallel and decoupled Lieb lattice planes. We take $c = d/2$, $\Omega_{11} = 0.0539$, $\Omega_{12} = -0.0247$, $\Omega_{22} = 0.0135$ and $\Omega_3 = 0.0956$.

\begin{figure}
\includegraphics[width=1.0\linewidth]{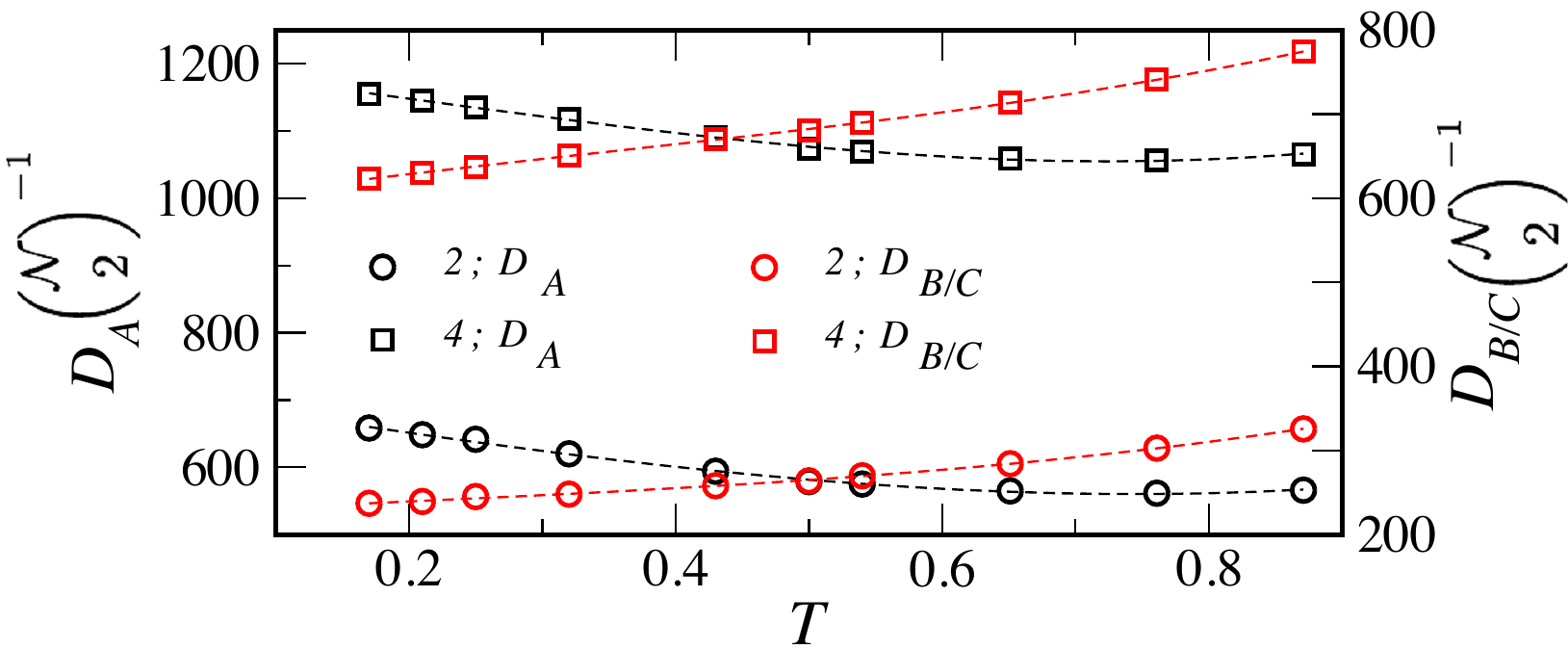}
\caption{\label{fig:three} Sublattice-resolved and spatially-averaged double occupancy $D_\alpha$~(\ref{eq:D_a}) vs. temperature $T$ for $U=2$ and $\mathcal{N}= 2,4$, in the presence of harmonic trapping. The chemical potential is fixed at $\mu = 0$ ($n = 3$ at the trap center for $\mathcal{N} = 2$, and $n = 6$ for $\mathcal{N} = 4$).}
\end{figure}

\begin{figure}
\includegraphics[width=1.0\linewidth]{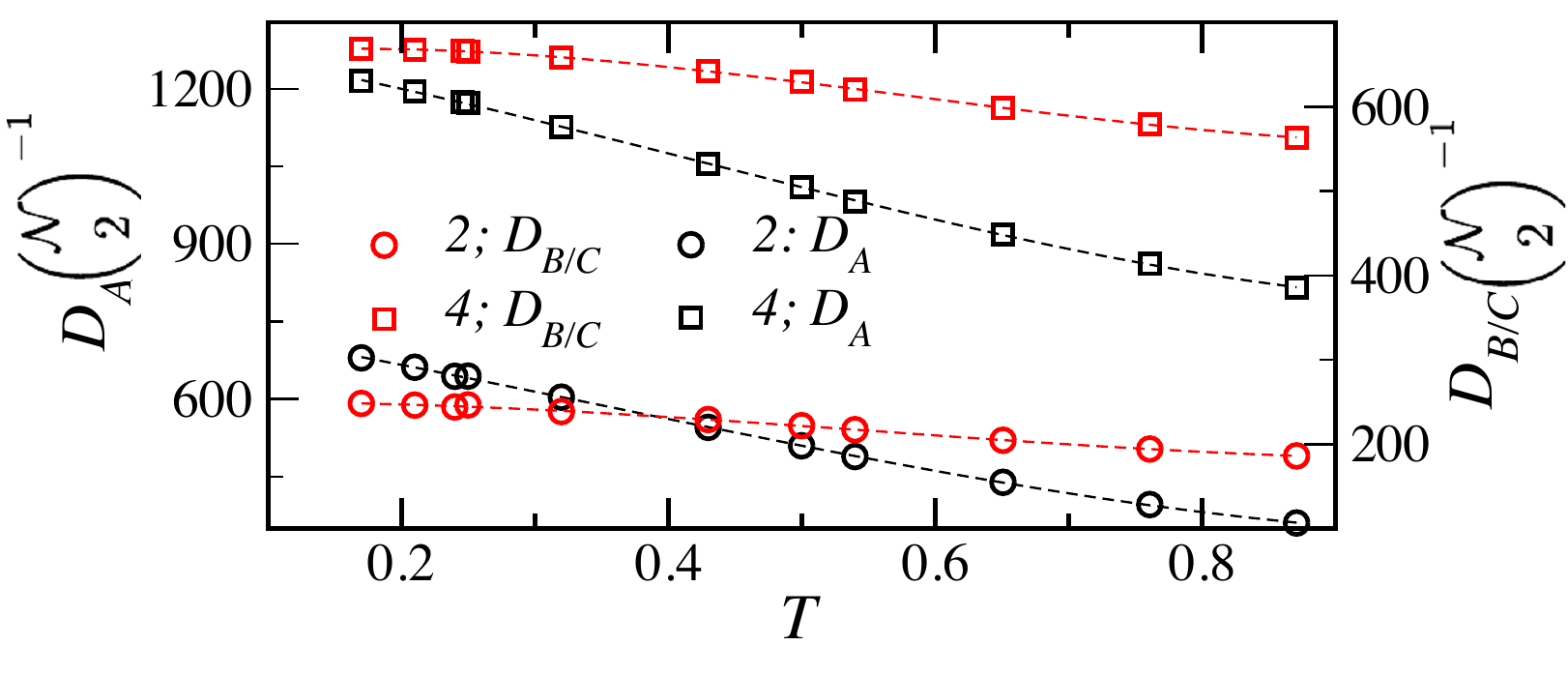}
\caption{\label{fig:four} Same as in Fig.~\ref{fig:three}, with the only difference that the total particle number $N$ is kept fixed, specifically $N = 1.2\times 10^4 (2.4\times 10^4)$ for $\mathcal{N} = 2(4)$ components.}
\end{figure}

The quantity $D_\alpha$ as a function of temperature is shown in Fig.~\ref{fig:three}. One can see that the different behaviour of the double occupancy in the two inequivalent sublattices remains visible even in the case of the spatially-averaged double occupancy. Indeed, $D_A$ decreases with increasing temperature almost up to the highest temperatures provided in Fig.~\ref{fig:three} for both $\mathcal{N} = 2$ and $4$. A change in the sign of $\partial_T D_A$ is visible around $T = 0.8$. On the other hand, the spatially-averaged double occupancy on the $B/C$ sublattices is monotonically increasing with temperature for any number of components. This is the same non-Fermi liquid behavior of the double occupancy in the thermodynamic limit shown for $\mathcal{N}=2$ in Fig.~2 in the main text.

In Fig.~\ref{fig:three} we consider the case of half-filling at the trap center, corresponding to the chemical potential fixed at $\mu = 0$. On the other hand it might be more convenient from an experimental point of view to fix the number of particles rather than the chemical potential. 
The sublattice-resolved and spatially-averaged double occupancy in the case of fixed particle number is shown in Fig.~\ref{fig:four}. In general for fixed particle number the local filling at the trap center is different from half-filling. The total particle number is chosen in a such a way that the filling is approximately $n = 3$ ($n = 6$) for $\mathcal{N} = 2$ ($\mathcal{N} = 4$) components at the intermediate temperature $T = 0.5$. 

In the case of fixed particle number (Fig.~\ref{fig:four}) the local density decreases at the trap center since the atomic cloud becomes more spread out with increasing temperature. This explains why even on the $B/C$ sublattices the double occupancy decreases with increasing temperature, if the particle number is fixed. This is simply a consequence of the fact that quite generally the double occupancy is a monotonically increasing function of the filling, as shown in Fig.~2 in the main text and in Fig.~\ref{fig:two} in the lower panel.
Even in the case of fixed particle number, the different behaviour of the double occupancy on the two inequivalent sublattices is 
still revealed by the different rates at which $D_\alpha$ decreases with temperature. As shown in Fig.~\ref{fig:four}, the rate is higher on the $A$ sublattice than on the $B/C$ sublattices, which is consistent with the results for fixed chemical potential shown in Fig~\ref{fig:three}.

\bibliography{biblio}